\documentclass[twocolumn,epsfig]{revtex4}
\usepackage{epsfig,color}
\begin{document}
\title{Effect of isospin dependence of cross section on symmetric and neutron rich systems}

\author{Aman D. Sood$^1$ }
\email{amandsood@gmail.com}
\address{
$^1$SUBATECH,
Laboratoire de Physique Subatomique et des
Technologies Associ\'ees \\University of Nantes - IN2P3/CNRS - Ecole des Mines
de Nantes 
4 rue Alfred Kastler, F-44072 Nantes, Cedex 03, France}
\date{\today}

\maketitle

\section*{Introduction}
The existing and upcoming
radioactive ion beam (RIB) facilities lead a way in exploring the role of isospin degree of freedom in both nuclear physics and astrophysics. The ultimate goal of such studies is to extract information on the
isospin dependence of in-medium nuclear effective interactions as
well as equation of state (EOS) of isospin asymmetric nuclear
matter. The later quantity especially the symmetry energy term is
important for both nuclear physics and astrophysics community. Heavy-ion collisions induced by neutron rich beams provide unique opportunities to investigate the
isospin-dependent properties of asymmetric nuclear matter,
particularly the density dependence of symmetry energy \cite{li08}. Experimentally symmetry energy is not a directly
measurable quantity and has to be extracted from observables
related to symmetry energy. It has also been shown in ref \cite{gautam2,gautam3} that the collective transverse in-plane flow and its disappearance is sensitive to the various isospin effect like Coulomb repulsion, symmetry energy, isospin dependence of cross section and there is a complex interplay among above mentioned reaction mechanisms. In refs. \cite{gautam2,gautam3}, it was also shown that the Coulomb repulsion dominates over the symmetry energy in isospin effects on collective flow. Sood \cite{sood4} studied the disappearance of flow i.e. balance energy $E_{bal}$ for isotopic series of Ca with N/Z varying from 1 to 2 i.e. $^{40}Ca+^{40}Ca$ to $^{60}Ca+^{60}Ca$ and found that N/Z dependence of $E_{bal}$ is sensitive not only to symmetry energy but its density dependence as well. Here we aim to explore the effect of isospin dependence of cross section on symmetric and neutron rich system. We also aim to explore whether the analysis is affacted if one discusses in terms of "$E_{bal}$ as a function of N/Z or N/A" of the system.

\section*{Results and Discussion}
Using IQMD model \cite{hart98} we have simulated several thousand events at incident energies around
balance energy in small steps of 10 MeV/nucleon for each
isotopic system of Ca+Ca having N/Z (N/A) varying from 1.0 to 2.0 (0.5-0.67) for
the semicentral colliding geometry range of 0.2 - 0.4. For the
transverse flow, we use the quantity "\textit{directed transverse
momentum $\langle p_{x}^{dir}\rangle$}" which is defined as
\cite{sood4}
\begin {equation}
\langle{p_{x}^{dir}}\rangle = \frac{1} {A}\sum_{i=1}^{A}{sign\{
{y(i)}\} p_{x}(i)},
\end {equation}
where $y(i)$ is the rapidity and $p_{x}$(i) is the momentum of
$i^{th}$ particle. The rapidity is defined as
\begin {equation}
Y(i)= \frac{1}{2}\ln\frac{{\textbf{{E}}}(i)+{{\textbf{p}}}_{z}(i)}
{{{\textbf{E}}}(i)-{{\textbf{p}}}_{z}(i)},
\end {equation}
where ${\textbf{E}}(i)$ and ${\textbf{p}_{z}}(i)$ are,
respectively, the energy and longitudinal momentum of $i^{th}$
particle.
In fig. 1(a) we display the E$_{bal}$ as a function of
N/Z of the system. Solid green circles represent the calculated
E$_{bal}$. Lines are the linear fit to E$_{bal}$. We see that
E$_{bal}$ follows a linear behavior $\propto$ m$\ast$N/Z. As the
N/Z of the system increases, the mass of the system increases due
to addition of neutron content. In addition, the effect of
symmetry energy also increases with increase in N/Z. To check the
relative contribution of increase in mass with N/Z and symmetry
energy towards the N/Z dependence of E$_{bal}$, we make the
strength of symmetry energy zero and calculate E$_{bal}$. The
results are displayed by open circles in fig. 1(a). E$_{bal}$
 again follows a linear behavior $\propto$ m$\ast$N/Z. However, E$_{bal}$ decreases very slightly with increase in
 N/Z, whereas when we include the symmetry energy
 also in our calculations then the $\mid m \mid$ increases by 3
 times which
 shows that N/Z dependence of E$_{bal}$ is highly sensitive
 to the symmetry energy. To explore the effect of isospin dependence of cross section (i.e. $\sigma_{nP}=3\sigma_{nn}=3\sigma_{pP}$), we switch off the symmetry energy and also make the cross
 section isospin independent (i.e. $\sigma_{np}$ = $\sigma_{nn}$
 and calculate E$_{bal}$ for two extreme N/Z. The results are displayed in
 fig. 1(a) by open squares. Again E$_{bal}$ follows a linear
 behavior. We see that the E$_{bal}$ for both $^{40}$Ca +
 $^{40}$Ca and $^{60}$Ca + $^{60}$Ca increases as expected.
 However, the increase in E$_{bal}$ for system with N/Z = 1 is
 more as compared to the system with N/Z = 2. This is  because
 with increase in N/Z the neutron number increases due to which
 neutron-neutron and neutron-proton collisions pairs increase.
 However, the increase in number of neutron-neutron collision
 pairs is much larger as compared to neutron-proton collision
 pairs. Therefore, the possibility of neutron-proton collision is
 much less in system with N/Z = 2. Therefore the effect of
 isospin dependence of cross section decreases with increase in
 N/Z.
\begin{figure}[!t] \centering
\vskip 0.5cm
\includegraphics[angle=0,width=6cm]{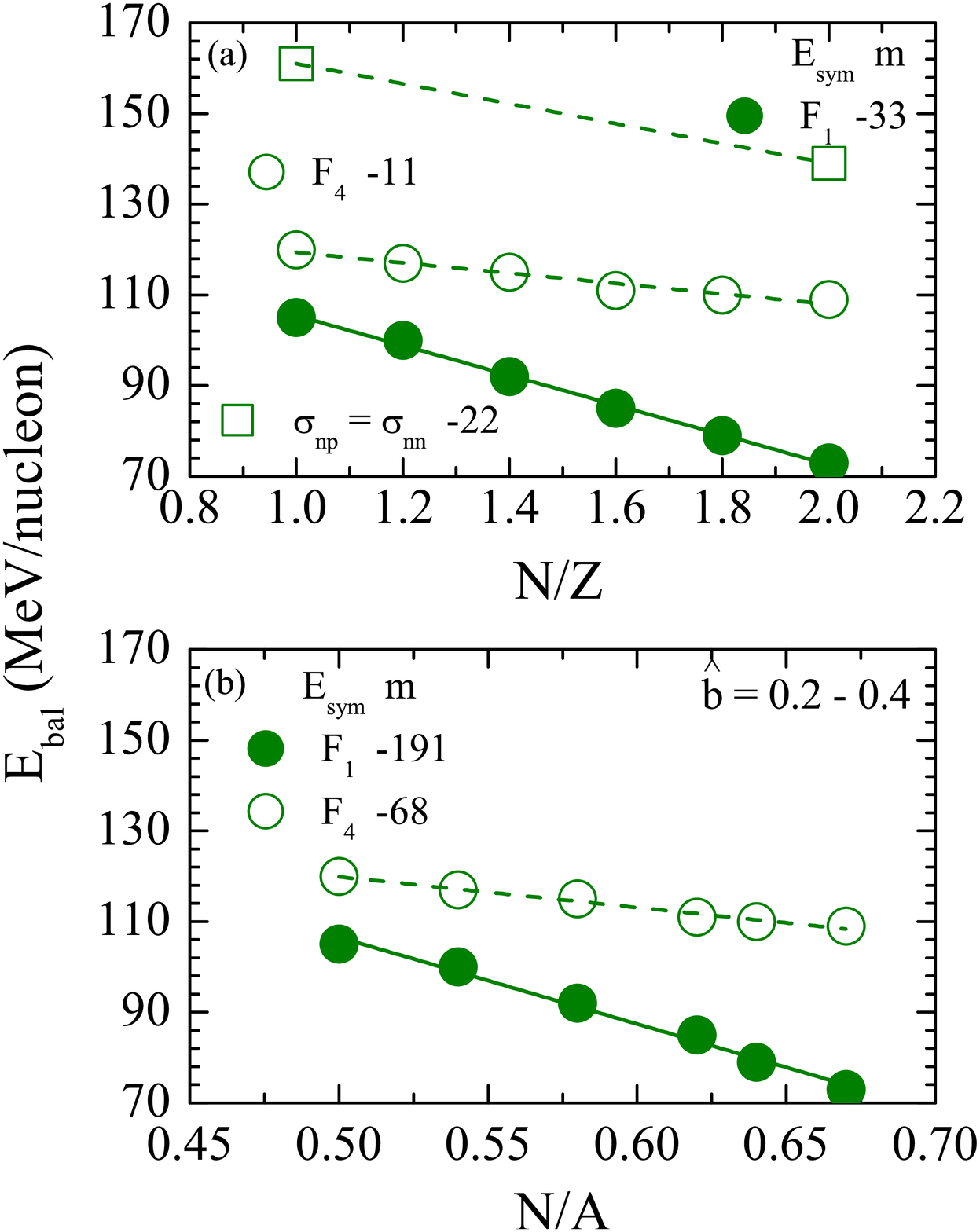}
\caption{\label{fig1} E$_{bal}$ as a function of N/Z (upper panel) and N/A (lower panel) of system
for E$_{sym} \propto F_{1} (u)$ and F$_{4}$. Lines are linear fit proportional to m. Various symbols are
explained in the text.}
\end{figure}

In fig. 1(b), we display the E$_{bal}$ as a function of N/A of the system. Symbols have same meaning as in fig. 1(a). Again E$_{bal}$ follows
a linear behaviour with m = -191 and -68, respectively, for F$_{1}$ (u) and F$_{4}$ where F$_{1}$ (u) represents symmetry energy $\propto$ $\frac{\rho}{\rho_{0}}$ and F$_{4}$ represent calculations without symmetry energy. However, the percentage difference $\Delta E_{bal}$ \% (where $\Delta E_{bal}$ \% =
$\frac{E_{bal}^{F_{1}(u)}-E_{bal}^{F_{4}}}{E_{bal}^{F_{1}(u)}}$) is same (about 65\%) in both the figs. 1(a) and 1(b) which
shows that the effect of symmetry energy is on N/Z dependence of $E_{bal}$ is same whether we discuss in terms of N/Z or N/A.

\section*{Acknowledgments}
This work has been supported by a grant from Indo-French Centre for the Promotion of Advanced Research (IFCPAR) under project no 4104-1.


\end{document}